\documentclass{aa}
\usepackage{graphics,epsf,epsfig,graphicx,rotating}
\begin{document}

\thesaurus{ 
		02.08.1 
		02.20.1 
		09.03.1; 
		09.07.1; 
		09.11.1; 
		09.19.1; 
             } 
\title{ N-body simulations of self-gravitating gas
in stationary fragmented state }
   
\author{B. Semelin and  F. Combes} 
\offprints{B. Semelin} 
\institute{
	Observatoire de Paris, DEMIRM, 61 Av. de l'Observatoire, 
	F-75014, Paris, France 
	} 
\date{Received XX XX, 1999; accepted XX XX, 2000} 
\authorrunning{Semelin \& Combes} 
\titlerunning{Simulations of self-gravitating gas}
\maketitle 
 
\begin{abstract}   
The interstellar medium is observed in a hierarchical fractal structure
over several orders of magnitude in scale. Aiming to understand
the origin of this structure, we carry out
numerical simulations of molecular cloud fragmentation,
taking into account self-gravity, dissipation and energy input.
Self-gravity is computed through a tree code,
with fully or quasi periodic boundary conditions. 
Energy dissipation is introduced
through cloud-cloud ineslatic collisions. Several schemes are tested for
the energy input. It appears that energy input from galactic shear
allows to achieve a stationary clumped state for the gas, avoiding
final collapse. When a stationary turbulent cascade is established,
it is possible to derive 
meaningful statistical studies on the data such as  
the fractal dimension of the mass distribution.
\end{abstract} 
  
\section{Introduction} 

From many studies over the last two decades, it has been established that the
interstellar medium (ISM) has a clumpy hierarchical structure, approaching  a
fractal structure independent of scale over 4 to 6 orders of magnitude in sizes
(Larson 1981, Scalo 1985, Falgarone et al 1992, Heithausen et al 1998). The 
structure extends up to giant molecular clouds (GMC) of
100 pc scale, and possibly down to 10 AU scale, as revealed by HI absorption VLBI
(Diamond et al 1989, Faison et al 1998) or extreme scattering events in quasar
monitoring (Fiedler et al. 1987, Fiedler et al. 1994).  It is not yet clear which
mechanism is the main responsible for this structure; it could be driven by
turbulence, since the Reynolds number is very high, or self-gravity, 
since clouds appear to be virialized on most of the scales,  with the help of 
magnetic fields, differential rotation, etc...

One problem of the ISM turbulence is that the relative velocities of clumps are
supersonic, leading to high dissipation, and short lifetime of the structures.
An energy source should then be provided to maintain the turbulence. It could be
provided by star formation (stellar winds, bipolar flows, supernovae, etc... 
Norman \& Silk 1980). However, the power-law relations
observed between size and line-width, for example, are the same in regions of
star-formation or quiescent regions, either in the galactic disk, or even
outside of the optical disk, in the large HI extensions, where very little star
formation occurs. In these last regions alternative processes of energy input 
must be available. A first possibilty is the injection by the galactic shear.
Additionally, since there is no heating sources in the
gas, the dense cold clumps should be bathing at least in the cosmological
background radiation, at a temperature of 3K (Pfenniger \& Combes 1994). In any
case, in a nearly isothermal regime, the ISM should fragment recursively (e.g.
Hoyle 1953), and be Jeans unstable at every scale, down to the smallest
fragments, where the cooling time becomes of the same order as the
collapse time, i.e. when a quasi-adiabatic regime is reached (Rees 1976).

In this work, we try to investigate the effect of self-gravity
through N-body simulations. We are not interested in star formation, but
essentially in the fractal structure formation, that could be driven essentially
by gravity (e.g. de Vega et al. 1996). Our aim is to reach a quasi-stationary
state, where there is statistical equilibrium between coalescence and
fragmentation of the clouds. This is possible when the cooling (energy
dissipated through cloud collisions, and subsequent radiation) is compensated by
an energy flux due to external sources~: cosmic rays,
star formation, differential shear...
Previous simulations of ISM fragmentation have been performed to
study the formation of condensed cores, some with isolated boundary conditions
(where the cloud globally collapses and forms stars, e.g. Boss 1997, Burkert et
al. 1997), or in periodic boundary conditions (Klessen 1997, Klessen et al.
1998). The latter authors 
assume that the cloud at large scale is stable, supported by
turbulence, or other processes. They follow the over-densities in a given
range of scales, and schematically stop the condensed cores as sink particles
when they should form structures below their resolution.
We also adopt periodic boundary conditions, since numerical simulations are very
limited in their scale dynamics, and we can only consider scales much smaller 
than the large-scale cut-off of the fractal structure. We are also limited by
our spatial resolution: the smallest scale considered is
 far larger than the physical small-scale cut-off
 of the structures. Our dynamics is computed within this range of scales.

Our goal is to
achieve long enough integration time for the system to reach a stationary state,
 stationary in a statistical sense. 
This state should have its energy confined in a narrow domain, thus presenting 
the 
density contrasts of a fragmented medium while avoiding gravitational collapse 
after a few dynamical times. Only when those conditions are fulfilled, can we
try to build a meaningful model of the gas. Let us consider for example the
turbulence concepts.   

Theoretical attempts have been made to describe the interstellar medium as a
turbulent system. While dealing with a system both self-gravitating and 
compressible, 
the standard approach has been to adapt Kolmogorov picture of incompressible
turbulence. The first classical assumption is that the rate of energy transfer 
between scales is constant within the so-called inertial range. This
inertial range is delimited by a dissipative scale range at small scales and
a large scale where energy is fed into the system. In the case 
of the interstellar medium, the energy source can be the galactic shear, or
the galactic magnetic field, or, on smaller scales, stellar winds and such.
In this classical picture of turbulence, one derives the relation 
$v_L \sim L^{1/3}$, where $v_L$ is the velocity of structures on scale $L$.
If we consider now that the structures are virialized at all scales, we get the
relation $\rho_L \sim L^{-4/3}$, where $\rho_L$ is the density of structures
on scale $L$. This produces a fractal dimension $D=1.66$. A more consistent
version is to take compressibility into account in Kolmogorov cascade; then
$\rho_L v_L^3 / L \sim cst$. Adding virialization, we get the relations
$v_L \sim L^{3/5}$ and $\rho_L \sim L^{-4/5}$. This produces the fractal
dimension $D=2.2$.
It should be emphasized again that all these scenarios assume a quasi-stationary
regime. 

However, numerical simulations of molecular cloud fragmentation
have been, so far, carried out
in dissipative schemes which allow for efficient clumping
(e.g Klessen et al 1998, Monaghan \& Lattanzio 1991). As such, they do not
reach stationary states. Our approach is to add an energy input making up
for the dissipative loss. After relaxation from initial conditions
we attempt to reach a fragmented but non-collapsing state of the medium that
does not decay into homogeneous or centrally condensed states. It is then 
meaningful to
compute the velocity and density fields power spectra and test the standard
theoretical assumptions. It is also an opportunity to investigate the
possible fractal structure of the gas. Indeed, starting from a weakly perturbed
homogeneous density field, the formation time of a fractal density field 
independent of the initial conditions is at 
the very least of the order of the free-fall time, and more likely many times longer.
A long integration time should
permit full apparition of a fractal mass distribution, independent of the
initial conditions.

This program encounters both standard and specific difficulties. Galaxies
clustering as well as ISM clumping require large density contrasts, that is
to say high spatial resolution in numerical simulations. This point is tackled, be it
uneasily, by adaptative-mesh algorithms or tree algorithms, or multi-scale schemes
($\hbox{P}^3\hbox{M}$). Another problem
is the need for higher time resolution in collapsed region. This is CPU-time
consuming or is dealt with by multiple time steps. This point is of primary
sensitivity in our simulation since we need to follow accurately the
internal dynamics of the clumps and filaments to avoid total energy divergence. 
Finally we need a long integration time to reach the stationary state. 
This is directly in competition for CPU-time with spatial resolution for given 
computing resources.

\section{ Numerical methods}

\subsection{ Self-gravity}

In our simulation we use N-body dynamics with a hierarchical tree algorithm
and multipole expansion to compute the forces, 
as designed by Barnes \& Hut (1986, 1989). The number of particles is 
between $10\;000$
and $120\;000$ particles. We are limited by the fact that we need long
integration time in our study.

Within the N-body framework the tree algorithm
is a $N\ln(N)$ search algorithm picking the closest neighbours for exact
force computation and sorting farther out particles in hierarchical boxes for
multipole expansion. The size of the boxes used in the expansion is set,
for the contribution of a given region to the force, by a control parameter
$\theta$
defining the maximal angular size of the boxes seen from the point where the
force is computed. Typical values used for $\theta$ are 0.5 to 1.0. Multipole 
expansion is carried out to quadrupole terms. 
As usual a short
distance softening is used in the contribution of the closest particle to the
interaction. The softening length is taken as $\sim 1/10$ of the 
inter-particle distance of the homogeneous state.

\subsection{Boundary conditions}
There are two ways to erase the finite size effect and avoid spherical 
collapse in a N-body simulation. We can choose either quasi-periodic boundary 
conditions or fully periodic boundary conditions following Ewald method
(Hernquist et al 1991). In quasi-periodic conditions the interaction
of two particles is computed between the first particle and
the closest of all the replicas of the other particle. The 
replicas are generated in the periodization of the simulation box to the whole
space. In the fully periodic conditions, the first particle interacts with all 
the replicas of the other one. The relevance of each method will be discussed
in section \ref{3dbc}. We will compare them for simple free-falls using
the fractal dimension as diagnosis.

\subsection{ Time integration and initial conditions}
 
The integration is carried out through a multiple time steps leap-frog
scheme. Making a Keplerian assumption about the orbits of the particles, we
have the following relation between time step and the local interparticle 
distance: $\delta\tau \sim (\delta l)^{3 \over 2}$. In a fractal medium,
the exponent could be different. We do not take this into account since it would
lead us to modify the integration scheme dynamically according to the fractal
properties of the system. As the tree sorts particles in cells of size
$2^n d_0$ according to the local density, we should use $(2 \sqrt{2})^n \delta
t_0$ time steps. For simplicity we are using $3^n \delta t_0$. Then,
$n$ different time steps allow us to follow the dynamics with the same
accuracy over regions with density contrasts as high as $2^{3(n-1)}$.
 
We will use several types of initial conditions.  
The usual power law for the density power spectrum will be used for the
free falls. It is usually 
implemented by using Zel'dovitch approximation which is valid before first
shell crossing for a pressureless perfect fluid. The validity range of
this approximation is difficult to assess in an N-body simulation. We
use a similar implementation that does not however extend to non-linear 
regimes but also gives density power spectra with the desired shapes.
A gaussian velocity field is chosen with a specified power spectrum 
$P_v(k)\sim k^{\alpha-2}$, then the particles
are positioned at the nodes of a grid and displaced according to the velocity
field with one time step. The resulting density fluctuations spectrum is
$P_{\rho}(k) \sim k^{\alpha}$ according to the matter conservation.

\subsection{ Gas physics}

 There can be several modelizations for the ISM: either it is
considered as a continuous and fluid medium and simulated through
gas hydrodynamics (with pressure, shocks, etc..)
or, given the highly clumped nature of the interstellar
medium, and its highly inhomogeneous structure
(density variations over 6 orders of magnitude), it can
be considered as a collection of dense fragments, that dissipate
their relative energy through collisions.
Our choice of particle dynamics assumes that each particle stands for a clump 
and neglects the interclump medium. 
Another reason is that we expect a fractal distribution
 of the masses. This means a non-analytical density field which is not to be
easily handled by hydrodynamical codes. 

The dissipation enters the dynamics through a gridless sticky particles 
collision scheme. The collision round is periodically computed every 1 to 10
time steps. The frequency of this collision round is one way to adjust the
strength of the dissipation. Since, as the precise collision schemes will reveal
, we adopt a statistical treatment of the dissipation, it is not necessary to
compute the collision round at each time step.
Two schemes are investigated. The first uses the tree search to
find a candidate to collision with a given particle, in a sphere whose radius 
$r_c$ is a fixed parameter. Inelastic collisions are then 
computed, dissipating energy but conserving linear momentum. In the second
scheme, each particle has a probability to collide proportional to the inverse
of the local mean collision time. If it passes the probability test, the
collision is computed with the most
suitable neighbour. In this second scheme, dissipation may occur even in 
low density regions; in the first it occurs only above a given density threshold.
In practice this implies that, in the first scheme, dissipation happens only at
small scales, while in the second, it happens on all scales but is still 
stronger at small scales. 

Special attention must be paid to the existence of a small scale cutoff in
the physics of the system. Indeed, at very small scale ( $\sim 20 $ AU ), the
gas is quasi-adiabatic. 
Its cooling time is then much longer than the isothermal
free fall time at the same scale. Moreover, the mean collision time between 
clumps at this scale, in a fractal medium, is much shorter than the
cooling time. As a result, collisions completely prevent the already slowed 
down processes of collapse and fragementation in the quasi-adiabatic gas: the 
fractal
structure is broken at the corresponding scale. To take these phenomena into
account in our simulation, we need to introduce a large-density cutoff. This is
achieved by computing elastic or super-elastic collisions at scale $< 0.5 r_c$, 
instead
of inelastic collisions. The practical implementation follows the same
procedure as for the first inelastic collision scheme. In the case of 
super-elastic collisions, the overall energy
balance is still negative at small scales (due to a volume effect), however
we do introduce a mechanism of energy injection at small scale. This choice
can be furthermore justified by physical considerations.

When there is no energy provided by star formation (for example in
the outer parts of galaxies, where gas extends radially much beyond
the stellar disk), the gas is only interacting with the intergalactic 
radiation field, and the cosmic background radiation. The latter 
provides a minimum temperature for the gas, and plays the role of
a thermostat (at the temperature of 2.76K at zero redshift). The
interaction between the background and the gas can only occur
at the smallest scale of the fragmentation of the medium,
corresponding to our small scale cutoff; the radiative processes 
involve hydrogen and other more heavy elements (Combes \& Pfenniger 1997).
To maintain the gas isothermal therefore requires some energy 
input at the smallest scale.

Several schemes for the energy input have been tried. Reinjection at small scale
through added thermal motion has proved unable to sustain the system in
any other state than an homogeneous one. Then, turning to the turbulent point
of view, we have tried reinjection at large scale. Two main methods have been
tested; reinjection through a large scale random force field, and reinjection
through the action of the galactic shear. Their effect will be discussed in
section 5.

Finally, the statistical properties of the system can be described in many 
different way. We will use the correlation fractal dimension as diagnosis.
Definition and example of application of this tool are given in the appendix.
\section{ Simple 3-D free falls}

\subsection{ The choice of the boundary conditions}
\label{3dbc}

A fractal structure such as observed in the ISM or in the galaxies distribution
obeys a {\sl statistical} translational invariance. 
If we run simulations with vacuum boundary condition, this invariance is
grossly broken. On the other hand the quasi-periodic or 
fully-periodic conditions restore this invariance to some extent. They have 
been used in cosmological simulations. Hernquist et al (1991) have compared 
the two models to analytical results and found that the fully-periodic model 
simulates 
more accurately the self-gravitating gas in an expanding universe. This is
in agreement with expectations, since uniform expansion is 
automatically included in a fully-periodic model, while it is not in a 
quasi-periodic one.
Klessen (1996,1998) applied Ewald method to simulations of the 
interstellar medium. This is not natural since 
expansion is necessary to validate Ewald method. Nevertheless one can 
argue that
the dynamics is not altered by the use of Ewald method in regions with
high density contrasts.

Hernquist's study was on the dynamics of the modes. We are more 
interested in the fractal properties.
As far as we are concerned the question is wether the same initial 
conditions produce the same fractal properties in a cosmological framework
(fully-periodic conditions) and in a ISM framework (quasi-periodic conditions).
We will try to answer this question in section \ref{3dff}. 

\subsection{ The choice of the initial conditions}

The power spectrum of the density fluctuations is  most commonly described as
a power law 
in cosmological simulations. Then, different exponents in the power law produce
different fractal dimensions. The comoslogist can hope to find the fractal 
dimension from observations and go back to the initial power spectrum. In the
case of the interstellar medium however, it is unlikely that the
fractal dimension is the result of initial conditions since the life-time
of the medium is much longer than the dynamical time of the structures. 
In this regard, all initial conditions should be erased and produce the same 
fractal
dimension. We will check whether this is the case for power law density 
spectra with different exponents. Simple free falls with such initial conditions
will allow the comparison between the two types of periodic boundary conditions 
and will allow us to compare between 3D and 2D models and different dissipation
schemes. 

\subsection{ Simulations and results }

\label{3dff}

\begin{figure*}
\begin{center}
\epsfig{width=13cm,file=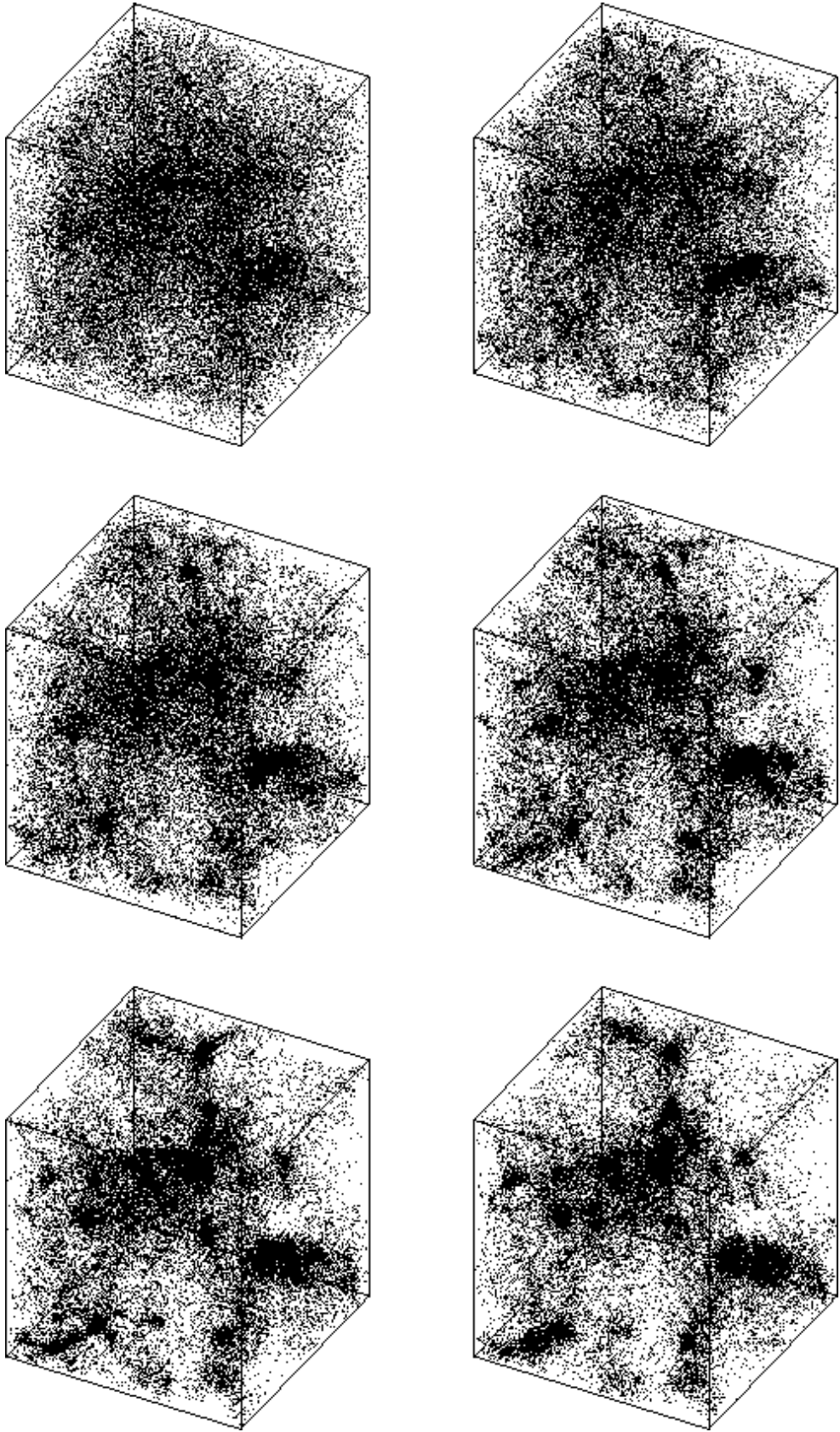}
\caption{\small Time evolution over about 0.5 free fall time.
The initial conditions are density 
fluctuations with power spectrum $P(k)\sim k^{-2}$. The simulation 
contains $117649$ $ (49^3)$ particles.}

\label{film}
\end{center}
\end{figure*}

The four first simulations are free falls from power law initial conditions in
either fully or quasi-periodic boundary conditions for two different power
law exponents: $\alpha=-1$ and $\alpha=-2$. About $120\,000$ particles were 
used in the simulations. The time evolution of the particle distribution for
$\alpha=-2$
and fully periodic condition is plotted in Fig. \ref{film}. If we compare
the qualitative aspect of the matter distribution with those found in 
SPH simulations (Klessen 1998), a difference appears. Filaments are less 
present in this N-body
simulation than in SPH simulations. We believe that this is due to the strongly
dissipative nature of SPH simulations. Indeed filaments form automaticaly,
{\sl even without gravity}, with $P(k) \sim k^{-2}$ initial conditions for the
density field.
Then, the strong dissipation of SPH codes is necessary to retain them;
$N$-body codes seem to show that gravity by itself is not enough.

\subsubsection{Fractal dimension}
\begin{figure*}
\begin{center}
\epsfig{width=13cm,file=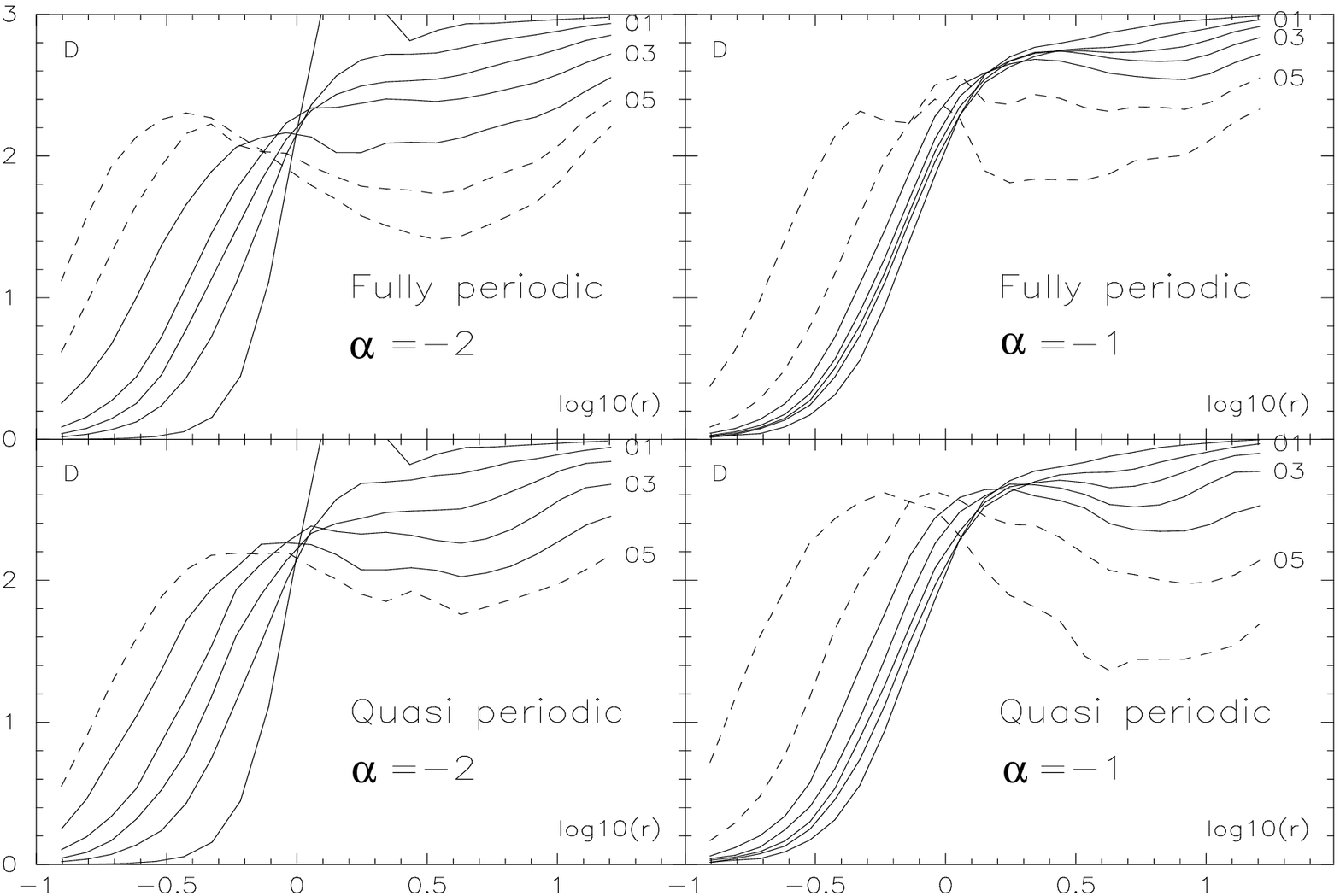}
\end{center}
\caption{\small The fractal dimension of the matter distribution
is plotted as a function of the log of the scale at different times of the 
free fall in four different conditions described under the curves. $\alpha$ 
designates the exponent of the power law of the density power spectrum.
Curves are labeled from early to late states.}
\label{fracdim}
\end{figure*}

For each of the four simulations, the fractal dimension as a function
of scale has been computed at different times of the free fall. 
Results are summarized in Fig.~\ref{fracdim}.

The fractal dimension of a mathematical fractal would appear either
as an horizontal straight line or as a curve 
oscillating around the fractal dimension if the fractal has 
no randomization (like a Cantor set). For our system, it appears that at 
sufficiently early stages of the free fall, and for scales above the 
dissipative cutoff, the matter
distribution is indeed fractal. However the dimension goes down with time
to reach values of 2 for $\alpha=-2$, and between 
$2$ and $2.5$ for $\alpha=-1$.
This is the fractal dimension for the total density, not for
the fluctuations only. 
Stabilization of the fractal dimension marginally happens before the dissipation
breaks the fractal.

One conclusion is however reasonable: as far as the fractal dimension is
concerned, the difference between fully-periodic boundary conditions and
quasi-periodic boundary conditions is not important. The deviation is a
bit stronger in the $\alpha=-1$ case, which is understandable since 
$\alpha=-1$ produces
weaker density contrasts in the initial conditions than $\alpha=-2$ and so the
dynamics is less decoupled from the expansion. The effect of this conclusion
is that we will not use the fully-periodic boundary conditions in further 
simulations since they require more CPU time and have no physical ground in 
the ISM framework.
Clump mass spectra are given in appendix B.

\section{Simple 2-D free falls}

We consider a 2-D system where particles have only two cartesian
space coordinates but still obey a ${1 \over r^2}$ interaction force.

\subsection{Motivations for the two dimensional study}

Turning to 2-D simulations allows access, for given computing resources,
to a wider scale-range of the dynamics. Or, for given computing resources
and a given scale-range, to larger integration times. The later is what we
will need in the simulation with energy injection. In practice we will choose
to use $10^4$ particles in 2-D simulations to access both a larger integration
time and a wider scale range. Indeed,
between a 3-D simulation with $N=10^5$ particles and a 2-D simulation with
$10^4$ particles one improves the scale range from
$[1, N^{1 \over 3}=46]$ to $[1,N^{ 1 \over 2}=100]$. At the same time one can increase
the integration-time by a factor of $\sim 10$ at equal CPU cost.

We can also add a physical reason to these computational justifications. The
simulations with an energy injection aim to describe the behaviour of
molecular clouds in the thin galatic disk. This medium has a strong anisotropy
between the two dimensions in the galactic plane and the third one. As such
it is a suitable candidate for a bidimensional modeling.

Moreover, studying a bidimensional system will allow us to check how the 
fractal dimension, dependent or not of the initial conditions, is affected 
by dimension of the space.

\subsection{ Simulations with 2 different dissipative schemes}

\begin{figure}
\begin{center}
\epsfig{width=7cm,file=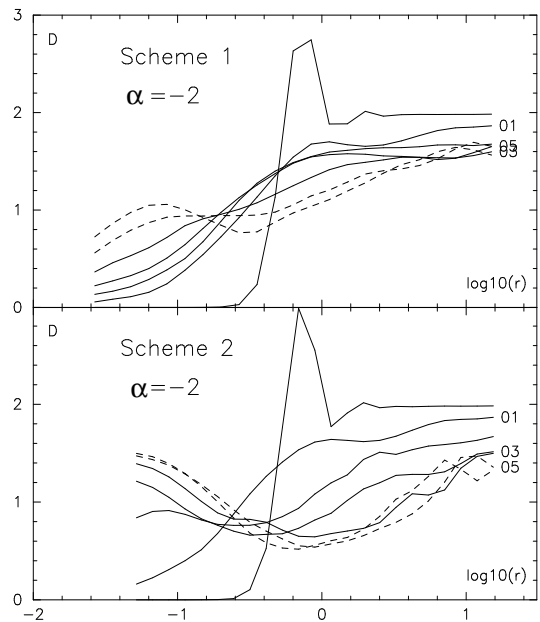}
\caption{ Fractal dimension as function of scale at different stages of the
time evolution. The simulations are two-dimensionnal with 10609 particles. The
dissipative scheme is modified between the two simulations according to
the description in sec 2.4 . }
\label{ff2d}
\end{center}
\end{figure}

Here again initial conditions with a power law
density power spectrum are used. Exponent $\alpha=-2$ is chosen to allow a 
comparison with the 3-D simulations. Boundary conditions are quasi-periodic. 
We have carried out two simulations with the two different dissipative schemes
described in section 2.4 .
The fractal dimensions at 
different stages of each of these two simulations are shown in Fig.~\ref{ff2d}.

In both simulations the fractal dimension of the homogeneous initial density
field is 2 at scales above mean inter-particle distance and 0 below. 
Then the density fluctuations
develop and the fractal dimension goes down to about $1.5-1.6$ in both 
simulations.  This value was not reached in 3-D simulations. This shows
that the dimension of the space has an influence on the fractal dimension of a 
self-gravitating system evolved from initial density fields with a power law 
power spectrum. Indeed spherical configurations are
favoured in 3 dimensions and may produce a different type of fractal than
cylindrical configurations which are favored in two dimensions.

Another point is that the fractal dimension of the system at small scales is
indeed sensitive to the dissipative scheme. The second scheme even alters
the fractality of large scales. So even if we believe that the fractality 
of a self-gravitating system is controlled by gravity, we must keep in mind 
the fact that
the dissipation has an influence on the result. In this regard, discrepancies
should appear between $N$-body simulations including SPH or other codes. 

The clump mass spectra are not conclusive due to
insufficient number of clumps formed with 10609 particles. 

\section{ 2-D simulations with an energy source}

As already stated, the ISM is a dissipative medium. Structures radiate their 
energy in a time-scale much shorter than the GMCs typical life time inferred
from observations. Therefore we are led to provide an energy source in the
system to obtain a life time longer than the free fall time.

\subsection{ The choice of the energy source}

Different types of sources are invoked to provide the necessary energy input:
stellar winds, shock waves, heating from a thermostat, galatic shear...
We have tested several possibilities. 

Our first idea was to model an interaction with a thermostat by adding
periodically a small random component to the velocity field. We have tried
to add it either to all the particles or only to the particle in the cool
regions. If enough energy is put into this random injection, the mono-clump
collapse is avoided. However the resulting final state is not fractal, 
nor does it show density structures on several different scales. It consists
in a few condensed points, like "blackholes", in a hot homogeneous phase.

The second idea is to model a generic large scale injection by introducing
a large scale random force field. We have tried stationary and fluctuating
fields. It is possible in this scheme, by tuning the input
and the dissipation, to lengthen the medium life time in an "interesting"
inhomogenous state by a factor 2 or 3. However the final state, after a
few dynamical times, is the same as for thermal input. 

We would like to emphasize that, in all the cases just mentioned, the existence
of super-elastic  collisions at very short distances, 
are {\em not at all} able to avoid the collapse of "blackholes". They are indeed beaten by inelastic
collisions which happen more frequently. Thus the density cutoff introduced
by superelastic collision, is not able all by itself to sustain or destroy
clumps.

Only one of the schemes we have tested avoids the blackhole/homogeneity duality.
In this scheme, energy is provided by the galactic shear. 
It produces an inhomogeneous state whose life time does not appear to be limited
by a final collapse in the simulations. We now describe this scheme in detail.

\subsection{ Modelisation of the galactic shear }

The idea is to consider the simulation box as a small part of a bigger
system in differential rotation. The dynamics
of such a sub-system has been studied in the field of planetary systems 
formation (Wisdom and Tremaine 1988).
Toomre (1981) and recently Hubert and Pfenniger (1999) have applied it to
galactic dynamics.

The simulation box is a rotating frame with angular speed $\Omega_0$ which is the
 galactic angular speed at the center of the box.
Coriolis force acts on the particles moving in the box. An additional external
force arises from the discrepancy between the inertial force and the local 
mean gravitational attraction of the galaxy. The balance between these two
forces is achieved only at points that are equally distant from the galactic
center as the center of the box. Let us consider a 2-dimensional case.
If $y$ is the orthoradial direction, and $x$ the radial direction 
the equations of motion are written:

\begin{eqnarray}
\ddot{y} &=& -2 \Omega_0 \dot{x} + F_y \nonumber\\
\ddot{x} &=& 2 \Omega_0 \dot{y} - 2 \Omega_0 r_0 
\left.{d \Omega \over d r}\right|_{r_0} \! x + F_x  \nonumber\\
 \nonumber\end{eqnarray}
The $F_{x,y}$ are the projections of the internal gravitational forces. The shear force
acting in the radial direction is able to inject into the system energy taken
from the rotation of the galaxy as a whole.

\begin{figure}
\begin{center}
\epsfig{width=6cm,file=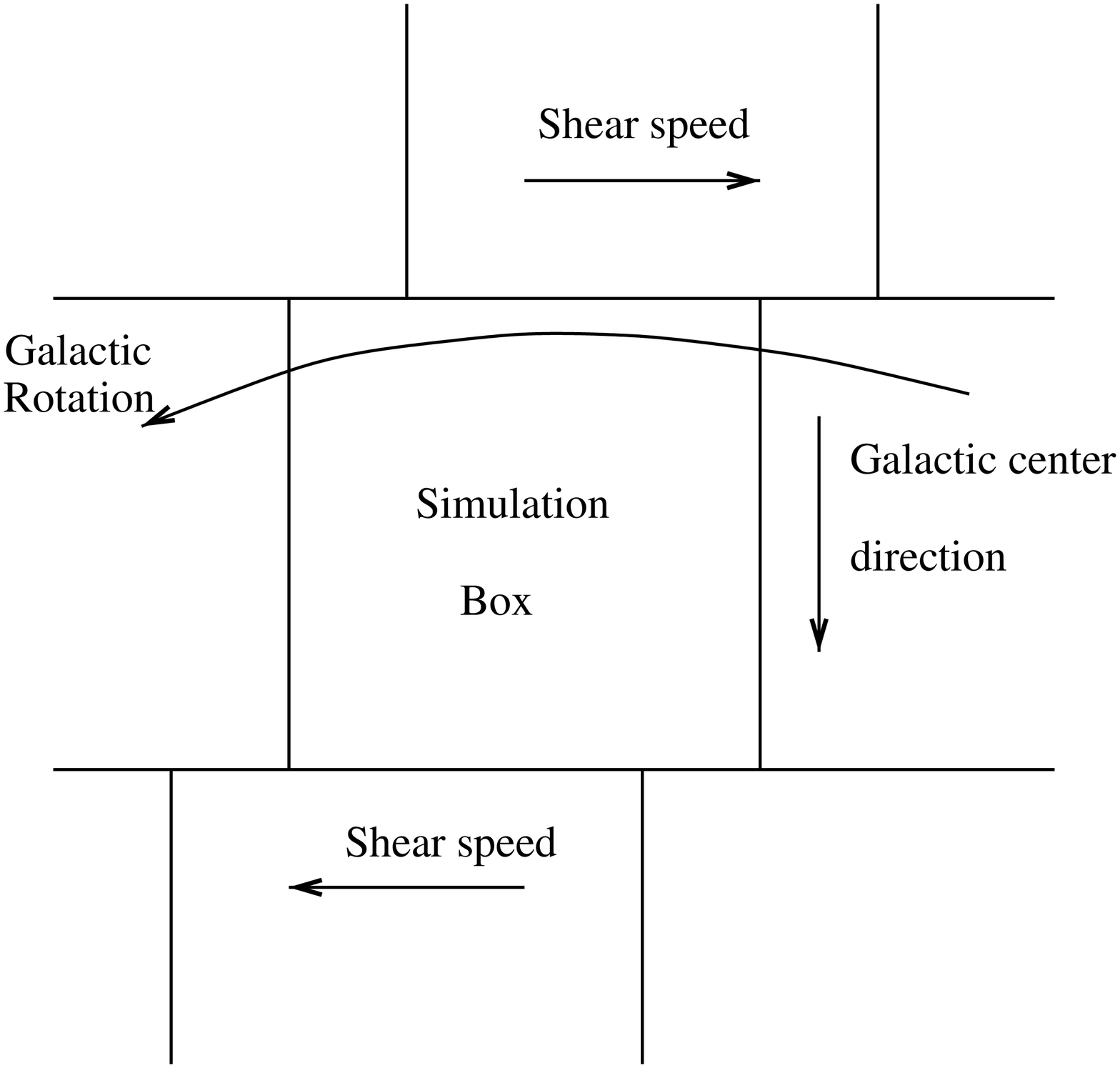}
\end{center}
\caption{ Effect of the galactic shear on the periodic conditions. The layers
of cells have a relative drifting speed. }
\label{shearsch}
\end{figure}

Some modifications must also be brought to the boundary condition. It is not
consistent with the differential rotation to keep a strict spatial periodicity.
To take differential rotation into account, layers of cells at different radii
must slide according to the variation of the galactic angular speed between them.
One particle still interacts with the closest of the replicas of another 
particle (including the particle itself). This is sketched on Fig. \ref{shearsch}.

\subsection{ Results}

\begin{figure*}
\begin{center}
\epsfig{width=11cm,file=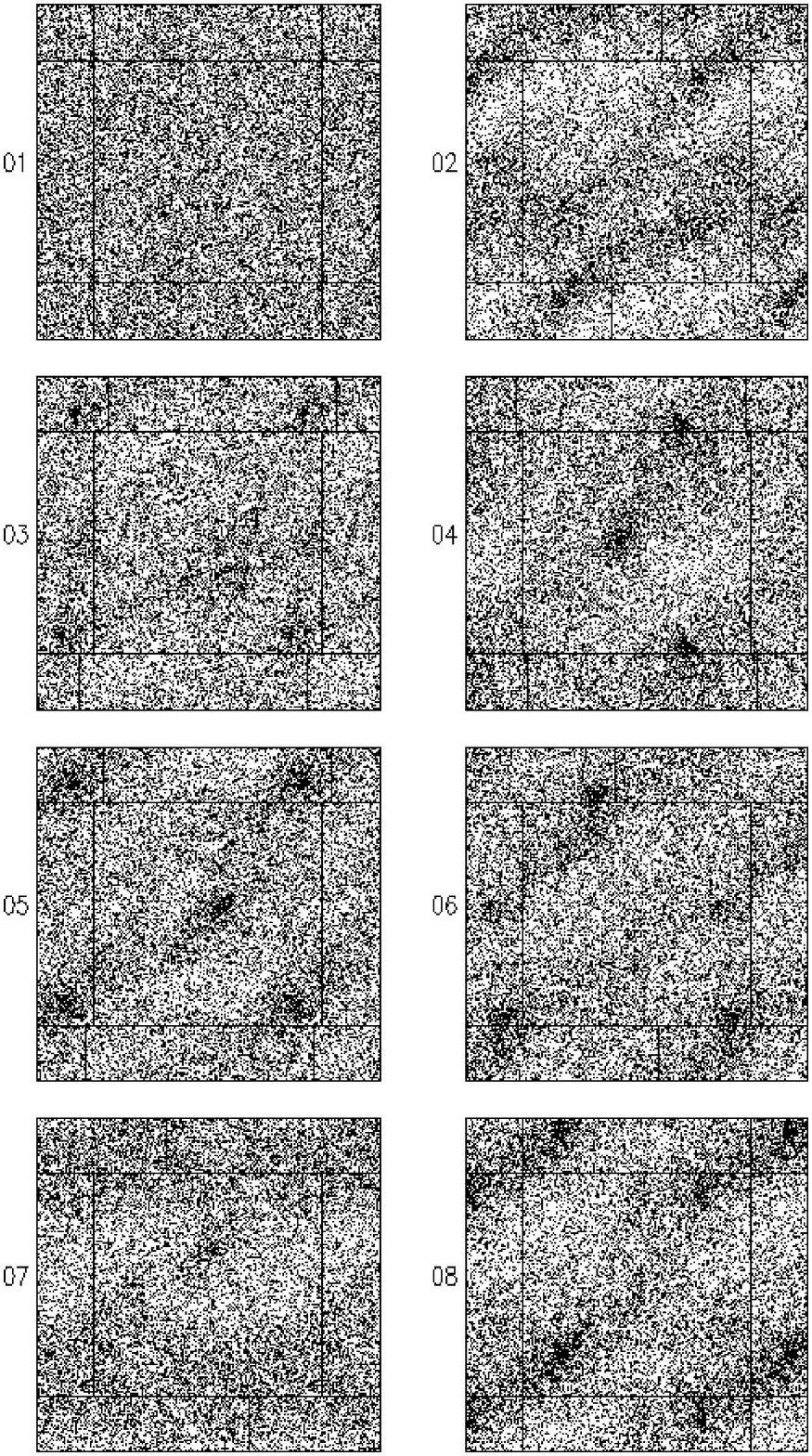}
\end{center}
\caption{2-D simulation with 10609 particles over 20 free fall times of the
total system. The different smapshots are taken at periodic time interval.
The galactic center is toward the bottom of the boxes. Sliding
replicas of the central box are shown for continuation}
\label{coriol}
\end{figure*}

We have computed simulations in two dimensions with 10609 particles. The 
direction orthogonal to the galactic disk is not taken into account. 
A tuning
between dissipation rate and shear strength is necessary to obtain a structured
 but non collapsing configuration.
However it is not a fine tuning at all. We found out that, in a range of values 
of the shear, the dissipation adapts itself to compensate the energy input so 
that we didn't encounter any limitation in the life time of the medium. 
The simulations have been carried out over more than 20 free-fall times. 
Snapshots from a simulation are plotted in Fig. \ref{coriol}. 

\begin{figure*}
\begin{center}
\epsfig{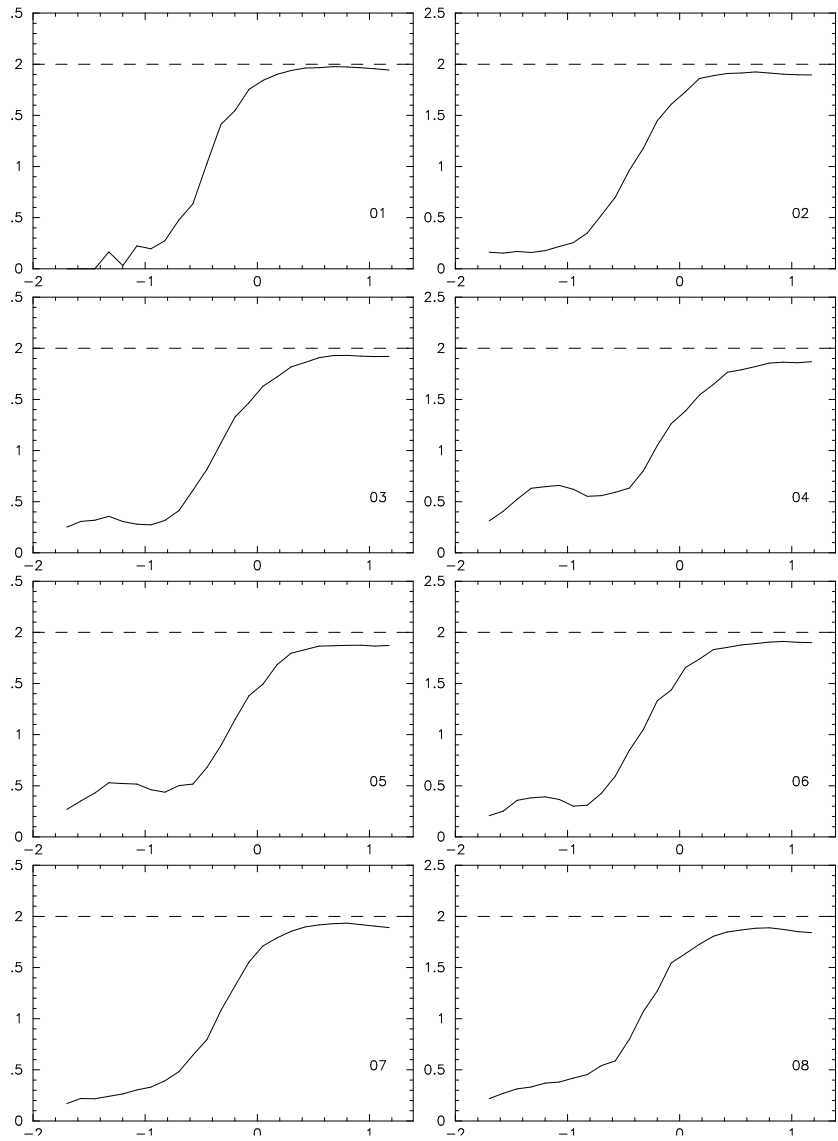}
\end{center}
\caption{ Fractal dimension computed at different epochs matching those of fig
\ref{coriol}.}
\label{df}
\end{figure*}

\subsubsection{Formation of persistent structures}
We can see that tilted stripes appear in the density field. This structure
appears also in simulations by Toomre (1981) and Hubert and Pfenniger (1999).
The new feature
is the intermittent appearance of dense clumps within the stripes. Theses
clumps remain for a few free fall times of the total system, then disappear,
torn apart by the shear. This phenomenon is unknown in the simple free fall 
simulations where a clump, once formed, gets denser and more 
massive with time. 
It also appeared that the tearing of the clumps by the
shear happens only if super-elastic collisions enact an efficient density 
cut-off. If we use elastic collisions only, below the dissipation scale,
thus enacting a less efficient density cut-off, the clumps persist. However
they do not collapse to the dense blackhole states encountered with different
large scale energy input schemes. So the combination of the galactic shear
and super-elastic collisions below the dissipation scale is necessary
to obtain destroyable clumps, while the shear alone produces rather 
stable clumps, and the super-elastic collisions alone cannot avoid complete
collapse.

While we cannot consider the density field as a fractal struture, 
it is indeed the
beginning of a hierarchical fragmentation since we have two levels of structures.
However the stripes are mainly the result of the shear. On the other hand the
clumps are the simple gravitational fragmentation of the stripes. These two
structures are not created by the exact same dynamical processes.

It must also be noted that the shear plays a direct role at large
scale (of the order of 10 pc), where the galactic tidal forces are comparable
to the self-gravity forces of the corresponding structures.   
But since the structures are fragmenting hierarchically, it is 
expected that its effect propagates and cascades down to the 
smallest scales.

\subsubsection{Fractal analysis}

The fractal dimension of the medium is plotted at different times in Fig.
\ref{df}. Two characteristic 
scales are clearly visible. At large scale, the scale of the stripes, the
dimension is about 1.9 with some fluctuations in time. At small scale, the
dimension fluctuates strongly due to the appearance and disappearance of clumps.
This shows through the appearance of a bump at the clump size. This size is 
about the dissipation scale. It is important to mention that the small scale 
range, while under the cutoff of the initial homogeneous condition, is not under
our dynamical resolution. The mean inter-particle distance
decreases from its initial value and is consistently followed by the 
$N$-body dynamics.

\subsubsection{Effect of the initial conditions}

We have already emphasized that in GMC simulations, the resulting structure 
should be independent of the initial conditions. We have performed the 
simulation for various values of the exponant of the power spectrum. In the
range $[-0.5,0.5]$ the long time behaviour is the one described above, thus
satisfying the required independence. However, for more steep spectrum value
like -2, the medium tend to collapse early into large clumps, thereby preventing
the shear to act efficiently.

We see that the behaviour is independent of the initial conditions as long as
the continuity of the medium at scale larger than the GMC size (the clump size
in the simulation) is not broken so that the shear can act to sustain the
GMC against collapse.

\section{Discussion and Conclusion}

We have shown that, in simulations without energy input, the evolution of
the fractal dimension depends on the initial conditions and also, to some
extent, on the dissipation schemes. Then we have investigated the effects of
an energy input on the dynamics of the system. This input is necessary to
prevent the total collapse of the system. Moreover, we do not want the
input to destroy the inhomogeneous fractal structure of the medium, as done by
 random
reinjections (thermal bath, random force field). Among
the solutions we have tested, only the highly regular force field provided by
the galactic shear preserves an inhomogeneous state. However this state is
quite strongly constrained by the geometry of the shear. Stripes appear as a result.
Interestingly, clumps are formed in these stripes and can be destroyed if we
have an efficient enough density cutoff . 

This state does not have a well-defined fractal dimension. Stripes are the effect
of a large scale action on the system and within our dynamic range in scale, we
cannot get rid of this boundary effect. If we had more resolution, the clumps
could fragment, and we might reach a scale domain free of boundary effects. 
One of our
prospects is to follow the inner dynamic of a clump with simulations in 
recursive sub-boxes going down in scale. This can produce a great dynamic scale
range but only in a small region of space. 

\begin{appendix}

\section{ Computing the fractal dimension}
 
By fractal dimension we mean more accurately the correlation fractal dimension.
We will use the notation $D_f$, defined as follows~: if we
consider a  fractal  set of points of equal weight located at $\hbox{\bf x}_j$,
the fractal dimension of the ensemble obeys the relation~:
 
$$
\lim_{r \rightarrow 0} r^{D_f}= \lim_{r \rightarrow 0} \left\langle \sum_j \int_
{|\hbox{\bf x}-\hbox{\bf x}_i|<r } \!\!\!\!
\delta(\hbox{\bf x}-\hbox{\bf x}_j)
d\hbox{\bf x} \right\rangle_i \; .
$$
The brackets stand for an ensemble average. In computational applications and
in physical systems, the $r \rightarrow 0$ limit is not reached. However the
relation should hold at scales where the boundary effects created by the finite
size of the system are negligible.
 
\begin{figure*}
\begin{center}
\epsfig{width=11cm,file=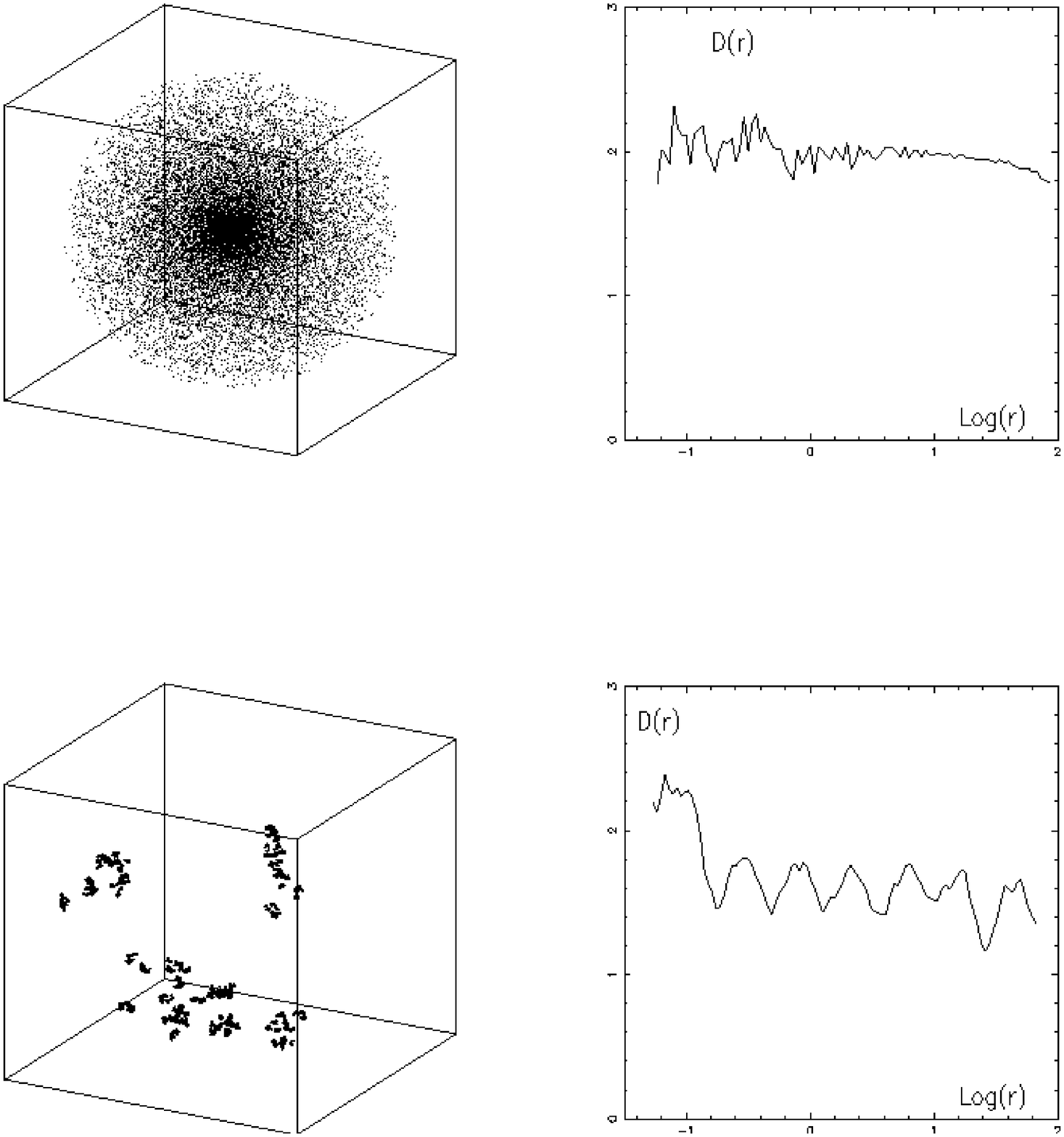}
\caption{\small The fractal correlation dimension is plotted as a function of
the scale for two different sets of points~: the first (not fractal) set is an
isothermal
sphere and a second is a fractal cantor-like set build recursively on 8 levels
with a small part of randomization. This shows that $D_f$ independent of $r$ is
neither a sufficient nor a mecessary condition of fractality}
\label{fractanal}
\end{center}
\end{figure*}

In practice, for a non-fractal ensemble of
points, $D_f$ usually depends on $r$ (but not always). For a fractal it is
independent of $r$, or at least it oscillates around a mean value. The latter
case can happen for a set with a strong scale periodicity like the Cantor set.
Examples are given in Fig. \ref{fractanal}.
As the above remarks show, $D_f$ is not a definitive criterion of
fractality.
 
According to the definition, the computation of the fractal dimension is based
on the following method.
$r^{D_f}$ is the mean mass in a sphere of radius $r$ centered on a particle.
This mass is actually computed for each particle using the tree search,
then it is averaged and the value $D_f(r)$ is derived. The average can be
made on a subset of points to lower the computation cost, the result is
then noisier.

\section{Clump mass spectrum}
 
Although the results are not very conclusive,
we give the mass distribution of the clumps in the 3-D simple freefall,
to allow comparison with observational data and other numerical simulations.
Plotting $\log_{10}(N)$
against $\log_{10}(m)$, the slope infered from observation data is $-0.5$. Thus
the clump mass spectrum behaves as a power law~: ${dN \over dm} \sim m^{-1.5}$.
 
To produce the mass spectrum, the first step is to define individual clumps.
We have used an algorithm very similar to the one proposed by Williams et al
(1994). We compute the density field from an interpolation of the
distribution of the particles with a cloud-in-cell scheme. It should be mentione
d that this
procedure tends to produce an artificially high number of very small clumps
associated to isolated particles or pair of particules which should not be
considered as clumps at all.
 
On Fig. \ref{cms} the clump mass spectrum is given at different times of the
evolution and for the two types of initial conditions ($\alpha=-1\,,\,-2$).
Quasi-periodic boundary
conditions are used. As time increases, as we can see, the clump mass spectrum
slope decreases to reach $-0.38 \pm 0.03$ for $\alpha=-1$ and $-0.18 \pm 0.03$
for $\alpha=-2$. At later times, the power law is broken by the final
dissipative collapse. According to this diagnosis, the case $\alpha=-1$ is
closer to the observed values. However, once again, this simulation provides
only a transient state which is unlikely to be a good model of a GMC.
 
\begin{figure}[t]
\begin{center}
\end{center}
\caption{\small Clump mass spectra for two values of $\alpha$ at different
evolution times. The given time-unit is about 1/10 of the free fall time.
At $t=5.$ the spectra show a $-0.38 \pm 0.03$ slope for $\alpha=-1$ and a
$-0.18 \pm 0.03$ for $\alpha=-2$. }
\label{cms}
\end{figure}
 
\end{appendix}

%

\end{document}